\documentclass[a4paper,aps,11pt,preprint,showpacs]{revtex4}
\usepackage{graphicx,amsmath,revsymb,bm,amssymb,amsfonts,amsthm}
\begin{document}

\title{Excitation of K-shell electrons by electron impact}

\author{A.V.~Nefiodov$^{\,\mathrm{a,b}}$ and G.~Plunien$^{\,\mathrm{a}}$}
\affiliation{$^{\mathrm{a}}$Institut f\"ur Theoretische Physik,
Technische Universit\"at Dresden, Mommsenstra{\ss}e 13, D-01062
Dresden, Germany \\
$^{\mathrm{b}}$Petersburg Nuclear Physics Institute, 188300
Gatchina, St.~Petersburg, Russia}

\date{Received \today}
\widetext
\begin{abstract}
The universal scaling behavior for the electron-impact excitation
cross sections of the $2s$ states of hydrogen- and helium-like
multicharged ions is deduced. The study is performed within the
framework of non-relativistic perturbation theory, taking into
account the one-photon exchange diagrams. Special emphasis is laid
on the near-threshold energy domain. The parametrical relationship
between the cross sections for excitation of multicharged ions with
different number of electrons is established.
\end{abstract}
\maketitle

{\bf 1.} Investigations of excitation processes of atomic ions by
electron impact are of fundamental importance. During last decades,
the problem has been intensively studied within the framework of
different sophisticated approaches (see the works \cite{1,2,3,4} and
references there). The deduction of the universal scaling behavior
of the differential and total cross sections, which allows one to
establish the generic features of the excitation processes, is of
particular interest. In this Letter, we solve the problem within the
framework of the consistent non-relativistic perturbation theory,
taking into account the one-photon exchange diagrams. Special
emphasis is laid on the near-threshold energy domain, because it
requires the correct treatment of the electron-electron and
electron-nucleus interactions.

{\bf 2.} Let us consider the inelastic scattering of an electron on
hydrogen-like ion in the ground state, which results in excitation
of a K-shell bound electron into the $2s$ state. We shall derive
formulae for differential and total cross sections of the process in
the leading order of non-relativistic perturbation theory with
respect to the electron-electron interaction. The nucleus of an ion
is treated as an external source of the Coulomb field. In a zeroth
approximation, the Coulomb functions are employed as electron wave
functions (Furry picture). The incident electron is characterized by
the energy $E=\bm{p}^2/(2 m)$ and the momentum $\bm{p}$ at
infinitely large distances from the nucleus, while the scattered
electron has the energy $E_1=\bm{p}_1^2/(2 m)$ and the asymptotic
momentum $\bm{p}_1$. The energy-conservation law reads $E=E_1 +
3I/4$, where $I=\eta^2/(2m)$ is the Coulomb potential for single
ionization from the K shell, $\eta =m \alpha Z$ is the average
mo\-men\-tum of the bound electron, $m$ is the electron mass, and
$\alpha$ is the fine-structure constant ($\hbar=1$, $c=1$).
Accordingly, the excitation process can occur, if $E \geqslant 3
I/4$. The parameter $\alpha Z$ is supposed to be sufficiently small
($\alpha Z \ll 1$), although we assume that the nuclear charge $Z
\gg 1$.

\begin{figure}[h]
\centerline{\includegraphics[scale=0.6]{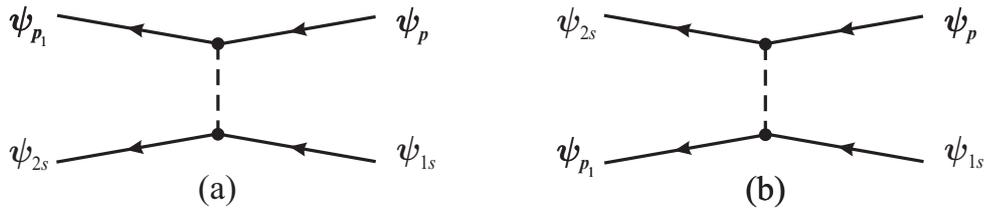}}
\caption{\label{fig1} Feynman diagrams for excitation of the K-shell
electron by electron impact. Solid lines denote electrons in the
external Coulomb field of the nucleus, while dashed line denotes the
electron-electron Coulomb interaction.}
\end{figure}

The process under consideration is described by the Feynman diagrams
depicted in Fig.~\ref{fig1}. In the initial and final continuum
states, the single-electron wave functions are denoted by
$\psi_{\bm{p}}$ and $\psi_{\bm{p}_1}$, respectively. Let us focus
first on the asymptotic non-relativistic energies $E$ within the
range $3I/4 \ll E \ll m$. In this case, $E_{1} \sim E$ and the
asymptotic momentum of the scattered electron is estimated as $p_1
\sim p \gg \eta$. Accordingly, one needs to take into account only
the Feynman diagram depicted in Fig.~\ref{fig1}(a). The contribution
of the exchange diagram turns out to be suppressed by the factor of
about $(\eta/p)^2$ and, therefore, can be neglected. Then the
amplitude of the process reads
\begin{equation}
{\mathcal A} = \int \langle \psi_{\bm{p}_1}| \bm{f}_1 \rangle
\langle \bm{f}_1 + \bm{f} | \psi_{\bm{p}} \rangle \frac{1}{f^2}
\langle \psi_{2s}| \bm{f}_2 \rangle \langle \bm{f}_2 - \bm{f}
|\psi_{1s} \rangle \frac{d\bm{f}}{(2\pi)^3}
\frac{d\bm{f}_1}{(2\pi)^3} \frac{d\bm{f}_2}{(2\pi)^3}  \,  .
\label{eq1}
\end{equation}
Note that the factor $4\pi\alpha$ originating from the photon
propagator is shifted from the amplitude to the formula for the
excitation cross section.

Since $p \sim p_1 \gg \eta$, the wave functions of both the incident
and scattered high-energy electrons can be approximated by the plane
waves (Born approximation). Integration over the intermediate
momenta in Eq.~\eqref{eq1} leads to
\begin{eqnarray}
{\mathcal A} &=&  N_{1s} \frac{1}{q^2} \Bigl( -
\frac{\partial}{\partial \lambda} \Bigr) \langle
\psi_{{2s}}|V_{i\lambda}| \bm{q} \rangle_{\left|\vphantom{\bigl(}\,
\lambda = \eta
\right. } \,   ,  \label{eq2} \\
\langle \bm{f}'| V_{i\lambda} | \bm{f}  \rangle &=& \frac{4\pi}{(
\bm{f}'- \bm{f} )^2 +\lambda^2} \,  ,
\end{eqnarray}
where $N^2_{1s}=\eta^3/\pi$ and $\bm{q}=\bm{p}- \bm{p}_1$ is the
momentum transfer. After taking the derivative with respect to
$\lambda$, one should set $\lambda=\eta$. Since the amplitude
\eqref{eq2} is a function of the square of the momentum transfer
$q^2$, it is convenient to introduce the dimensionless quantity
$x=(q/\eta)^2$. Accordingly, the amplitude ${\mathcal A}$ can be
cast into the form
\begin{eqnarray}
{\mathcal A} &=& N_{1s}N_{2s} \frac{1}{\eta^5}\, M
\,  , \label{eq4}\\
M &=& \frac{2^4 \pi}{(x + 9/4)^3} \,  , \label{eq5}
\end{eqnarray}
where $N^2_{2s}=\eta_2^3/\pi$ and $\eta_2=\eta/2$.

The differential cross section for the $1s$-$2s$ excitation is
related to the amplitude \eqref{eq4} via
\begin{equation}
d\sigma^*_{2s} = (4\pi \alpha)^2\frac{2\pi}{v} |{\cal A}|^2
\frac{d^3 \bm{p}_1}{(2\pi)^3} \, \delta(E - E_{1} - 3I/4) \,  .
\label{eq6}                                              
\end{equation}
Here $v=p/m$ is the absolute value of velocity of the incident
electron. Equation \eqref{eq6} determines the energy and angular
distributions of scattered electrons. The element of phase volume
for electrons scattered into the solid angle $d\Omega_1$ can be
written as
\begin{equation}
d^3\bm{p}_1 =  m p_1 \, dE_{1}\, d\Omega_1= \pi \frac{m}{p}\, dE_{1}
dq^2 \,  . \label{eq7}
\end{equation}
Integrating Eq.~\eqref{eq6} over the energy $E_{1}$ yields
\begin{equation}
d\sigma^*_{2s} =\frac{\sigma_0}{Z^4} \frac{M^2}{2 \pi^2
\varepsilon}\, dx \,   .
\label{eq8}                                          
\end{equation}
Here $\sigma_0=\pi a_0^2=87.974$ Mb, where $a_0=1/(m\alpha)$ is the
Bohr radius. The dimensionless function $M$ is given by the
expression \eqref{eq5}. In Eq.~\eqref{eq8}, we have also introduced
the dimensionless energy $\varepsilon =E/I$ of the incident
electron. The energy-conservation law implies $\varepsilon =
\varepsilon_1 + 3/4$, where $\varepsilon_1=E_1/I$ denotes the
dimensionless energy of the scattered electron.

To obtain the total cross section for the excitation of a K-shell
electron into the $2s$ state, the expression \eqref{eq8} should be
integrated over the variable $x$ within the range from
$x_1=(p-p_1)^2/\eta^2$ to $x_2=(p+p_1)^2/\eta^2$. This integration
can be performed analytically \cite{5}. The leading high-energy
contribution appears, while extending the integration from $x_1 =0$
to $x_2=\infty$. In this case, one obtains
\begin{eqnarray}
\sigma^*_{2s} &=& \frac{\sigma_0}{Z^4} \, Q(\varepsilon)   \,   ,
\label{eq9}       \\                               
Q(\varepsilon)&=& \frac{2^{7}}{\varepsilon} \int_{0}^{\infty}
\frac{dx}{(x + 9/4)^6}= \frac{2^{17}}{3^{10} 5 \varepsilon}  \, .
\label{eq10}                                        
\end{eqnarray}
The quantity $Z^{4}\sigma^*_{2s}$ is a universal function of the
dimensionless energy $\varepsilon$ of the incident electron. The
universal scaling $Q(\varepsilon)$ does not account for the exchange
effects and has a typical behavior for the excitation of states with
the zeroth matrix element of the dipole transition \cite{6}.
Equation \eqref{eq10} holds true only within the asymptotic
non-relativistic range $3/4 \ll \varepsilon \ll 2 (\alpha Z)^{-2}$.

Within the near-threshold domain, both the initial and final
electron momenta $p$ and $p_1$ are of the order of a characteristic
atomic momentum $\eta$. Correspondingly, the process occurs at
atomic distances of the order of the K-shell radius. This implies
that the Coulomb wave functions for both discrete and continuous
spectra should be employed already in zeroth approximation. In this
case, the direct and exchange Feynman diagrams depicted in
Fig.~\ref{fig1} are expected to give comparable contributions to the
excitation cross section.

Since the general method for evaluation of the Coulomb matrix
elements has been already described in details in Ref.~\cite{7}, we
present here the explicit expression for the amplitude $\mathcal{A}
=\sum\limits_{\beta} \mathcal{A}_{\beta}$ without derivation. The
individual contributions of both diagrams in Figs.~\ref{fig1}(a) and
(b) read
\begin{eqnarray}
\mathcal{A}_{\mathrm{a}} &=&
N_{1s}N_{2s}N_{p}N_{p_1}\dfrac{1}{\eta^5}\,M_{\mathrm{a}}\,
\delta_{\tau_1' \tau_1^{\vphantom{'}}}
\delta_{\tau_2' \tau_2^{\vphantom{'}}}  \,  , \label{eq11} \\
\mathcal{A}_{\mathrm{b}} &=&
N_{1s}N_{2s}N_{p}N_{p_1}\dfrac{1}{\eta^5}\,M_{\mathrm{b}}\,
\delta_{\tau_2' \tau_1^{\vphantom{'}}}
\delta_{\tau_1' \tau_2^{\vphantom{'}}}   \,  , \label{eq12} \\
M_{\mathrm{a}}&=& \frac{8\pi}{9}\, \xi^{3} \frac{1}{2\pi i}
\oint_{\gamma} \frac{dx}{x}
\Bigl(\frac{-x}{1-x}\Bigr)^{i\xi}\biggl\{\xi
\frac{\partial^{2}\Phi_{\mu}}{\partial\mu^{2}} - \frac{2}{3}
\frac{\partial\Phi_{\mu}}{\partial\mu}\biggr\}_{\left|
\vphantom{\bigl(}\, \mu = 3\xi/2 \right. }
\,  ,  \label{eq13}                           \\
M_{\mathrm{b}}&=& \frac{2}{\pi}\, \xi^{4}\Gamma_{\zeta}
\frac{\partial^2}{\partial \zeta
\partial \lambda}\int_{0}^{\infty} dy \int d\bm{\nu}
\,\Phi(\bm{y})\Phi_{1}(\bm{y}){}_{\left|{\lambda=\xi\hphantom{/2}}
\atop {\zeta =\xi/2} \right.}
\,  ,  \label{eq14}                              \\
\Phi_{\mu}&=&\frac{\left((\bm{n}_1 \varkappa - \bm{n}(1-x))^{2} - (x
+ i \mu)^{2}\right)^{i\xi_1 - 1}}{\left((1-x)^2 -(\varkappa + x +
i \mu)^{2}\right)^{i\xi_1} } \,  ,  \\
\Phi(\bm{y}) &=&  \frac{\left((\bm{n} - \bm{y})^2
+\zeta^2\right)^{i \xi -1} }{\left(y^2 - (1 +i \zeta)^2\right)^{i\xi}}
\,   ,   \label{eq16} \\
\Phi_{1}(\bm{y}) &=& \frac{\left((\bm{n}_1 \varkappa - \bm{y})^2
+\lambda^2\right)^{i \xi_1 -1} }{\left(y^2 - (\varkappa +i
\lambda)^2\right)^{i\xi_1}}  \,   ,  \label{eq17} \\
\Gamma_{\zeta}&=& 1 + \frac{\xi}{2}\frac{\partial}{\partial \zeta}\,
,  \quad \varkappa=\sqrt{1 - 3/(4 \varepsilon)} \,   ,  \\
\bm{n} &=& \frac{\bm{p}}{p}\,  ,  \quad \bm{n}_1 =
\frac{\bm{p}_1}{p_1} \,   ,  \quad \bm{\nu} = \frac{\bm{y}}{y} \,   , \\
\xi &=&\frac{1}{\sqrt{\varepsilon}}\,  ,  \quad
\xi_1=\frac{1}{\sqrt{\varepsilon_1}}=\frac{1}{\sqrt{\varepsilon - 3/4}}
\,  , \label{eq20}\\
N^2_{p} &=&\frac{2\pi\xi} {1 - e^{-2\pi \xi}} \,  , \quad N^2_{p_1}
= \frac{2\pi\xi_1} {1 - e^{-2\pi \xi_1}} \,  .   \label{eq21}
\end{eqnarray}
In Eq.~\eqref{eq13}, the integration contour $\gamma$ is a closed
curve encircling counter-clockwise the points 0 and 1. After taking
the derivatives in Eq.~\eqref{eq13} with respect to the parameter
$\mu$, one should set $\mu=3\xi/2$. In Eq.~\eqref{eq14}, after
taking the derivatives over $\lambda$ and $\zeta$, the parameters
should be set equal to $\lambda=\xi$ and $\zeta =\xi/2$,
respectively. The quantities $\tau_{1,2}^{\vphantom{'}}$ and
$\tau_{1,2}'$ denote the spin projections of the Pauli spinors in
the initial and final states, respectively.

The formulae \eqref{eq11}--\eqref{eq21} define the amplitude
$\mathcal{A} =\sum\limits_{\beta} \mathcal{A}_{\beta}$ of the
electron-impact excitation, in which the initial and final electrons
are characterized by definite polarizations. However, one is usually
interested in the processes, in which the initial particles are not
polarized, while polarizations of particles in the final state are
not measured. Accordingly, the differential cross section
\eqref{eq6} should be averaged over the polarizations of the initial
electrons and summed over the polarizations of the final electrons.
This can be achieved by means of the following substitution
\begin{equation}
|{\cal A}|^2 \to \overline{\vphantom{{\cal A}^2} | {\cal
A}|\,}\!{}^2 = \frac{1}{4} \sum_{\tau_{1}^{\vphantom{'}}, \tau_{1}'}
\sum_{\tau_{2}^{\vphantom{'}}, \tau_{2}'} |{\cal A}|^2 \, ,
\label{eq22}                                              
\end{equation}
where summations are performed over the electron polarizations in
both the initial and final states. The angular distribution of the
scattered electrons is given by
\begin{eqnarray}
\frac{d\sigma^*_{2s}}{d\Omega_1}&=& \frac{\sigma_0}{Z^4} \,
F(\varepsilon,\theta)\,   , \label{eq23} \\
F(\varepsilon,\theta) &=& \frac{2}{\pi}\frac{ \xi^{2}\, T}{ (1 -
e^{-2\pi \xi})( 1 - e^{-2\pi \xi_1})}\,  ,  \label{eq24}\\
T &=& \frac{1}{4}|M_{+}|^{2} + \frac{3}{4} |M_{-}|^{2} \,   .
\label{eq25}
\end{eqnarray}
Here the dimensionless functions $M_{\pm} = M_{\mathrm{a}} \pm
M_{\mathrm{b}}$, where particular contributions of the direct and
exchange diagrams are given by Eqs.~\eqref{eq13} and \eqref{eq14},
respectively. Due to the azimuthal symmetry of the problem, the
solid angle $d\Omega_1$ is just $d\Omega_1 = 2\pi \sin \theta
d\theta$, where $\theta$ is the angle between the vectors $\bm{n}$
and $\bm{n}_1$.

\begin{figure}[h]
\centerline{\includegraphics[scale=1.]{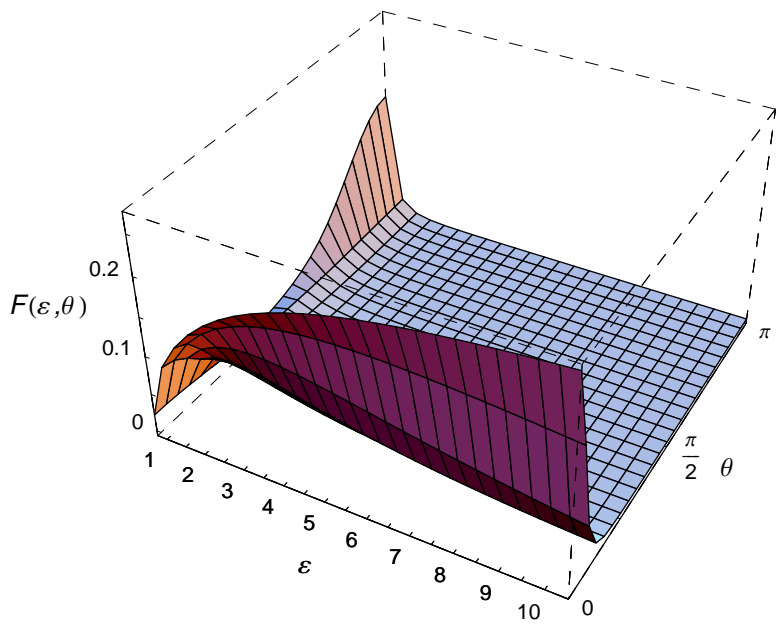}}
\caption{\label{fig2} The universal function $F(\varepsilon,\theta)$
is calculated according to Eqs.~\eqref{eq24} and \eqref{eq25} with
respect to the dimensionless energy $\varepsilon$ of the incident
electron and the scattering angle $\theta$.}
\end{figure}

\begin{figure}[h]
\centerline{\includegraphics[scale=1.]{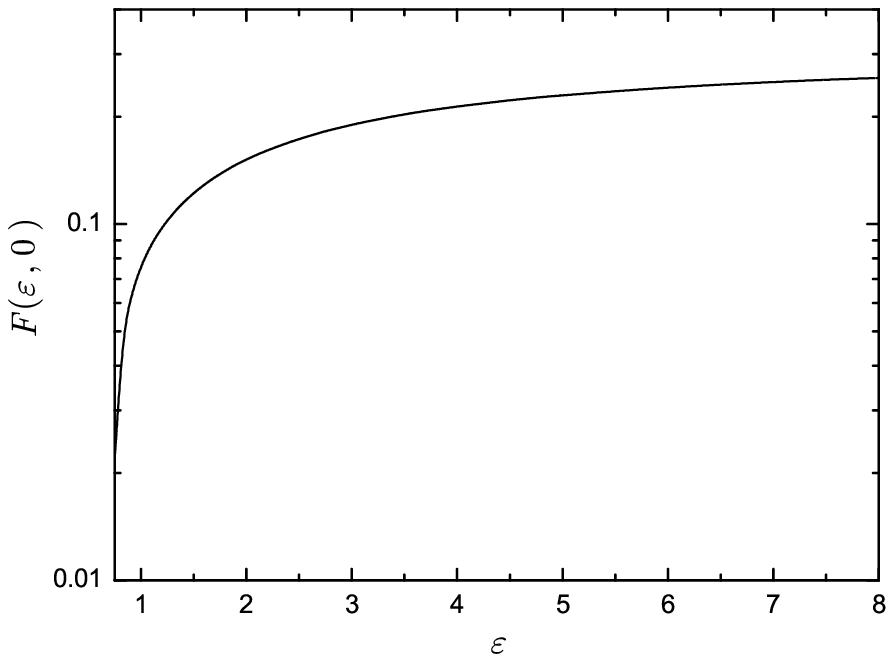}}
\caption{\label{fig3} The universal function $F(\varepsilon,\theta)$
is calculated according to Eqs.~\eqref{eq24} and \eqref{eq25} for
the particular scattering angle $\theta = 0$ (forward scattering).}
\end{figure}

\begin{figure}[h]
\centerline{\includegraphics[scale=1.]{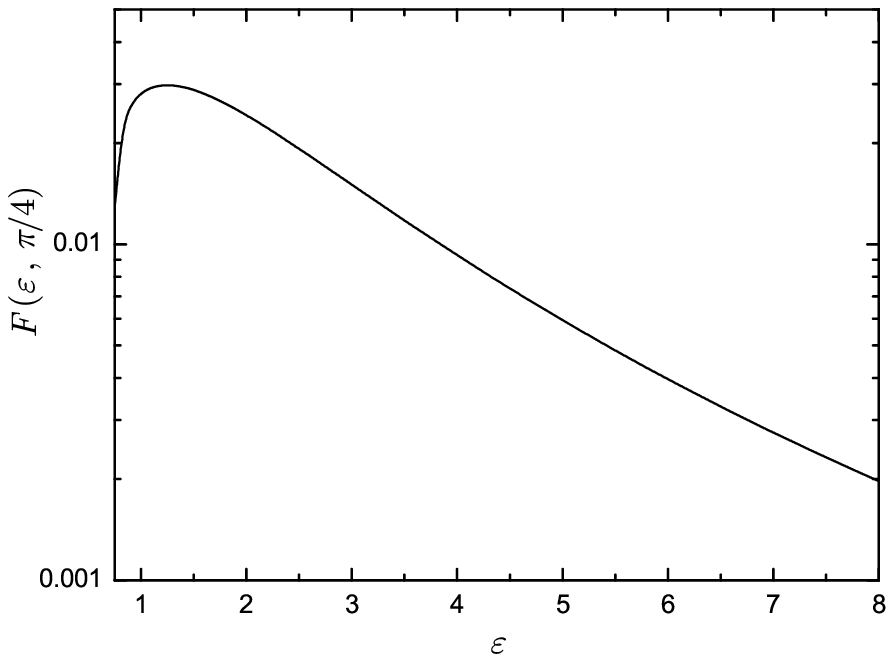}}
\caption{\label{fig4} The universal function $F(\varepsilon,\theta)$
is calculated according to Eqs.~\eqref{eq24} and \eqref{eq25} for
the particular scattering angle $\theta = \pi/4$.}
\end{figure}

\begin{figure}[h]
\centerline{\includegraphics[scale=1.]{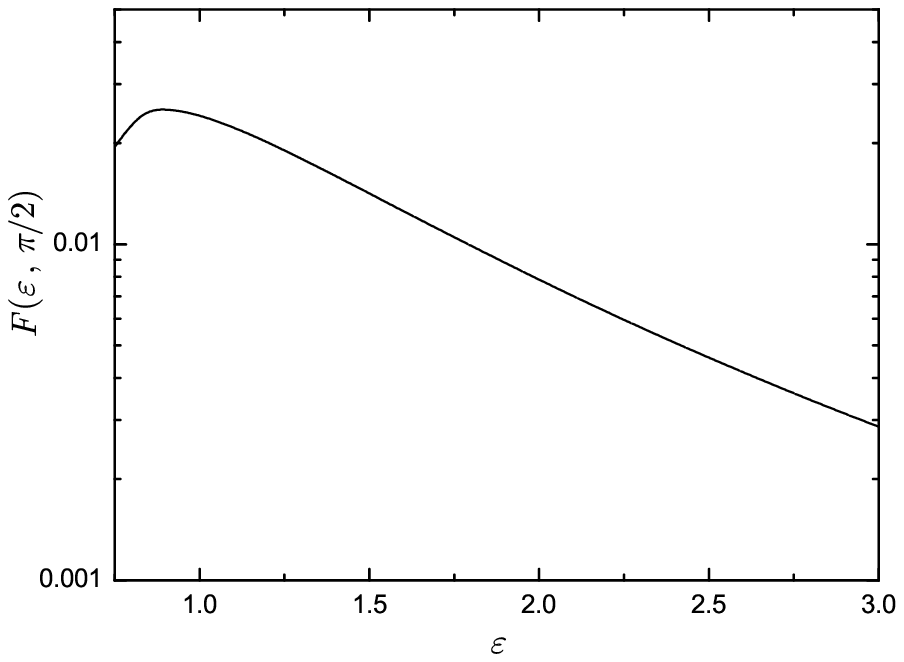}}
\caption{\label{fig5} The universal function $F(\varepsilon,\theta)$
is calculated according to Eqs.~\eqref{eq24} and \eqref{eq25} for
the particular scattering angle $\theta = \pi/2$.}
\end{figure}

\begin{figure}[h]
\centerline{\includegraphics[scale=1.]{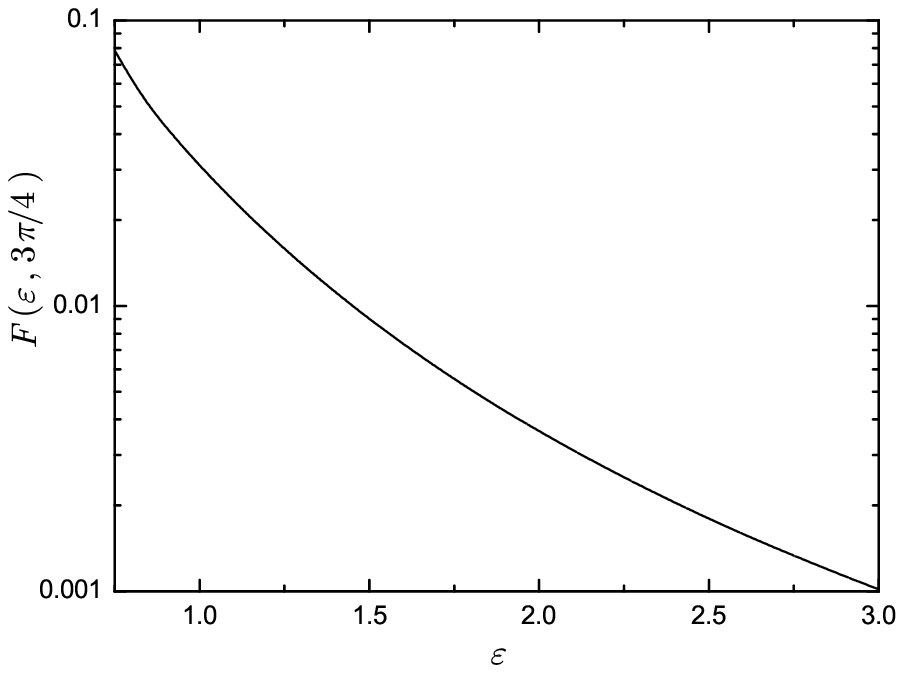}}
\caption{\label{fig6} The universal function $F(\varepsilon,\theta)$
is calculated according to Eqs.~\eqref{eq24} and \eqref{eq25} for
the particular scattering angle $\theta = 3\pi/4$.}
\end{figure}

\begin{figure}[h]
\centerline{\includegraphics[scale=1.]{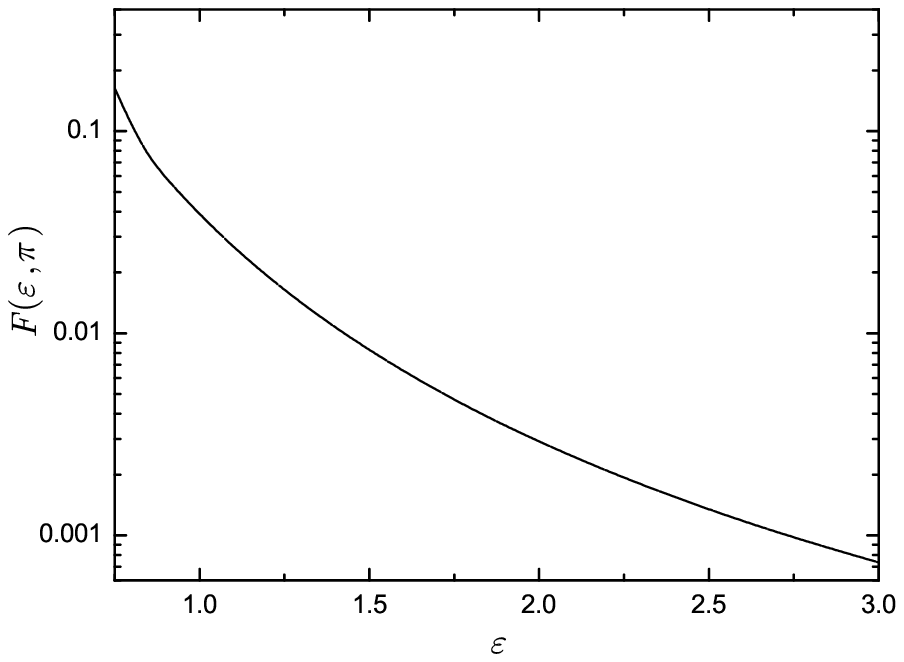}}
\caption{\label{fig7} The universal function $F(\varepsilon,\theta)$
is calculated according to Eqs.~\eqref{eq24} and \eqref{eq25} for
the particular scattering angle $\theta = \pi$ (backward
scattering).}
\end{figure}

The two-parameter function $F(\varepsilon,\theta)$, which is
universal for the non-relativistic energies $\varepsilon$ of the
incident electron within the range $3/4 \leqslant \varepsilon \ll 2
(\alpha Z)^{-2}$ and for any scattering angles $0 \leqslant \theta
\leqslant \pi$, is depicted in Fig.~\ref{fig2} within the
near-threshold domain. In Figs.~\ref{fig3}--\ref{fig7}, we present
also the energy behavior of the function $F(\varepsilon,\theta)$ for
a few particular angles $\theta=0$, $\pi/4$, $\pi/2$, $3\pi/4$, and
$\pi$, respectively. As it is seen, for energies $3/4 \leqslant
\varepsilon \lesssim 1$ the backward scattering is more preferable
rather than the forward scattering. In this case, the correct
account for both the direct and exchange diagrams appears to be
extremely crucial. However, with increasing energy $\varepsilon$,
the situation changes rapidly to the opposite: the forward
scattering becomes much more pronounced, while the backward
scattering turns out to be negligible. At high energies $\varepsilon
\gg 1$, the dominant contribution arises from the direct diagram
only.

\begin{figure}[h]
\centerline{\includegraphics[scale=1.]{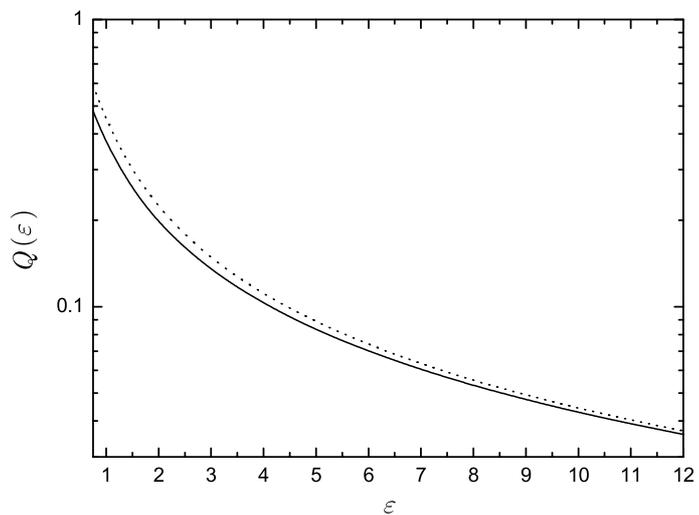}}
\caption{\label{fig8} The universal scaling $Q(\varepsilon)$ is
presented as a function of the dimensionless energy $\varepsilon$ of
the incident electron. Dotted line, plane-wave approximation
according to Eq.~\eqref{eq10}; solid line, exact calculation
according to Eqs.~\eqref{eq24}, \eqref{eq25}, and \eqref{eq27}.}
\end{figure}

Integrating Eq.~\eqref{eq23} over the solid angle yields the total
cross section
\begin{eqnarray}
\sigma^*_{2s} &=& \frac{\sigma_0}{Z^4} Q(\varepsilon) \,  ,
\label{eq26}  \\
Q(\varepsilon) &=& 2\pi \int_0^\pi \! F(\varepsilon, \theta) \sin
\theta d\theta \,  ,
\label{eq27}                                           
\end{eqnarray}
where $F(\varepsilon, \theta)$ is given by Eq.~\eqref{eq24}. The
universal function $Q(\varepsilon)$, which describes the whole
family of hydrogen-like targets with moderate nuclear charges $Z$,
is depicted in Fig.~\ref{fig8}. For comparison, we also draw there
the high-energy scaling \eqref{eq10} obtained within the framework
of the Born approximation. Although the latter is not applicable
within the near-threshold energy domain, the plane-wave results
appear to be in reasonable agreement with the exact predictions. For
example, at the threshold energy $\varepsilon_{\mathrm{th}}=3/4$,
Eq.~\eqref{eq27} yields $Q(\varepsilon_{\mathrm{th}})=0.482$, while
according to Eq.~\eqref{eq10} one receives just
$Q(\varepsilon_{\mathrm{th}})=2^{19}/(3^{11} 5)=0.592$.

It should be noted that the behavior of the universal function
$Q(\varepsilon)$ calculated according to Eq.~\eqref{eq27} is
relatively close to that presented in the work \cite{4}. However,
the numerical calculations performed in Ref.~\cite{4} for different
hydrogen-like ions within the framework of sophisticated methods
exhibit a slight dependence on the nuclear charge even for the
moderate values of $Z$ .

Just above the threshold, the $1s$-$2s$ excitation cross section
measured for the He$^+$ ion was found to be $\sigma^*_{2s}/\sigma_0
= 0.0128$ \cite{8}. Our prediction according to Eq.~\eqref{eq26} is
equal to $\sigma^*_{2s}/ \sigma_0 = 0.03$. The significant deviation
of these results seems to be caused by the correlation corrections
due to two-photon exchange diagrams, which have been neglected in
the present consideration.

{\bf 3.} Now we shall study the inelastic electron scattering on
helium-like ion in the ground state followed by the excitation of a
K-shell electron into the $2s$ state. To leading order of the
non-relativistic perturbation theory with respect to the
electron-electron interaction, one needs to consider only the
Feynman diagrams with one-photon exchange. In this case, it is
sufficient to take into account the interaction between two active
electrons, which participate in the excitation process. The
interaction with the second electron of target (spectator) is
neglected, since it first contributes only in the next-to-leading
order of the perturbation theory. Accordingly, the problem can be
reduced to that studied in the previous paragraph.

First, we shall obtain the cross section for impact excitation of
helium-like ion into the $1s2s$ configuration, when the energy terms
with different spin multiplicities cannot be resolved
experimentally. In this case, taking into account the number of
target electrons, the cross sections for helium- and hydrogen-like
ions are related as follows
\begin{equation}
d\sigma^*(\mathrm{He}) = 2 \, d\sigma^*_{2s} \,  , \label{eq28}
\end{equation}
where $d\sigma^*_{2s}$ is given by Eqs.~\eqref{eq23}--\eqref{eq25}.
Integrating Eq.~\eqref{eq28} over the solid angle $d\Omega_1$ yields
a similar relation for the total cross sections.

\begin{figure}[h]
\centerline{\includegraphics[scale=1.]{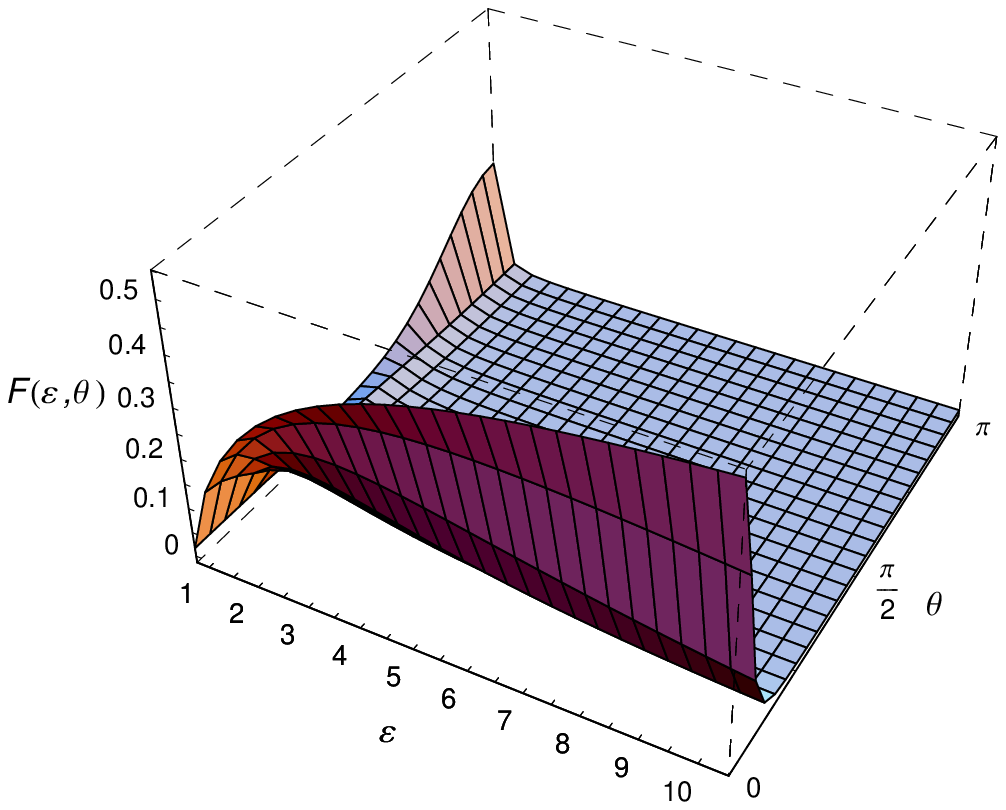}}
\caption{\label{fig9} The universal function $F(\varepsilon,\theta)$
describing the excitation of the singlet $2^1S$ state of helium-like
ion. The calculation is performed according to Eqs.~\eqref{eq24} and
\eqref{eq29} with respect to the dimensionless energy $\varepsilon$
of the incident electron and the scattering angle $\theta$.}
\end{figure}

\begin{figure}[h]
\centerline{\includegraphics[scale=1.]{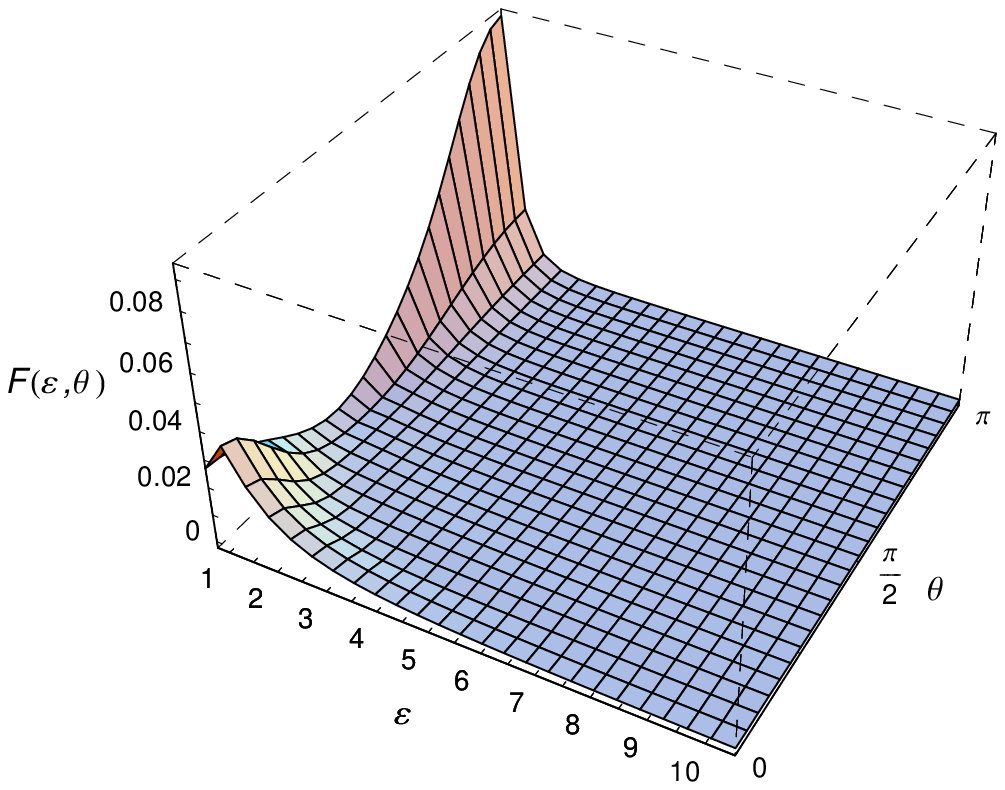}}
\caption{\label{fig10} The universal function
$F(\varepsilon,\theta)$ describing the excitation of the triplet
$2^3S$ state of helium-like ion. The calculation is performed
according to Eqs.~\eqref{eq24} and \eqref{eq29} with respect to the
dimensionless energy $\varepsilon$ of the incident electron and the
scattering angle $\theta$.}
\end{figure}

\begin{figure}[h]
\centerline{\includegraphics[scale=1.]{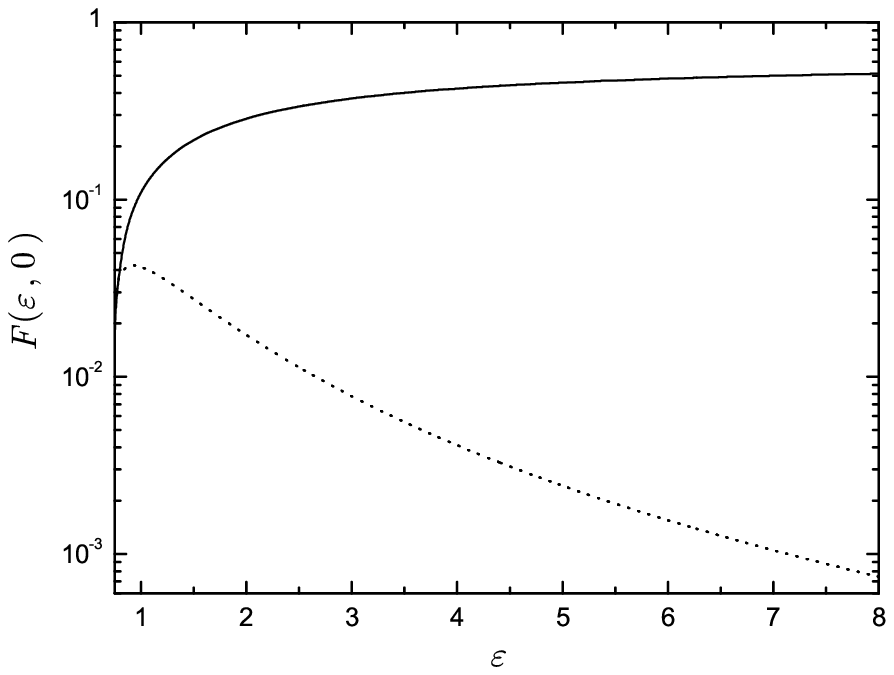}}
\caption{\label{fig11} The universal function
$F(\varepsilon,\theta)$ is calculated according to Eqs.~\eqref{eq24}
and \eqref{eq29} for the particular scattering angle $\theta = 0$
(forward scattering). Solid line, excitation into the singlet $2^1S$
state; dotted line, excitation into the triplet $2^3S$ state.}
\end{figure}

\begin{figure}[h]
\centerline{\includegraphics[scale=1.]{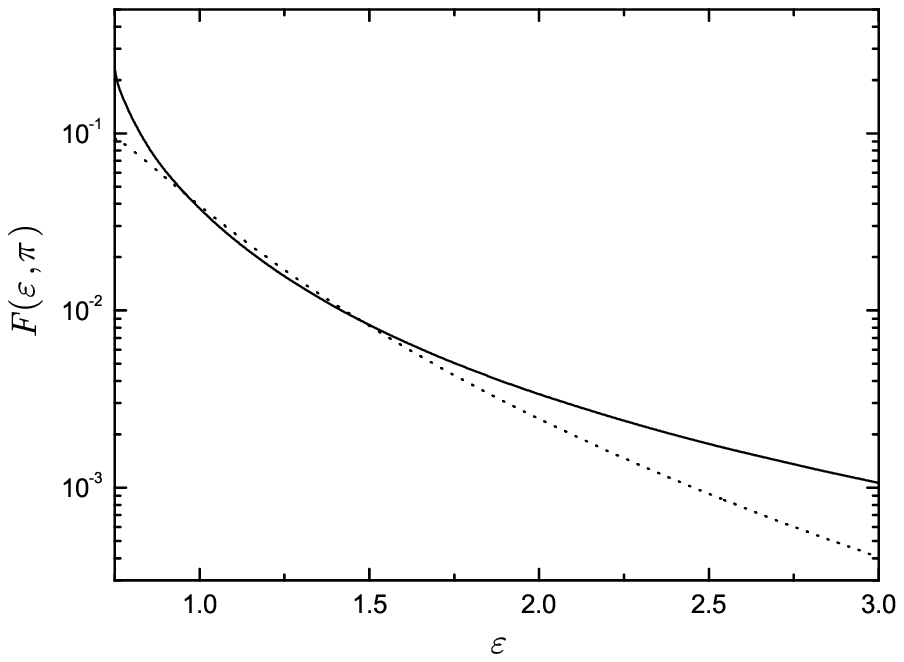}}
\caption{\label{fig12} The universal function
$F(\varepsilon,\theta)$ is calculated according to Eqs.~\eqref{eq24}
and \eqref{eq29} for the particular scattering angle $\theta = \pi$
(backward scattering). Solid line, excitation into the singlet
$2^1S$ state; dotted line, excitation into the triplet $2^3S$
state.}
\end{figure}

\begin{figure}[h]
\centerline{\includegraphics[scale=1.]{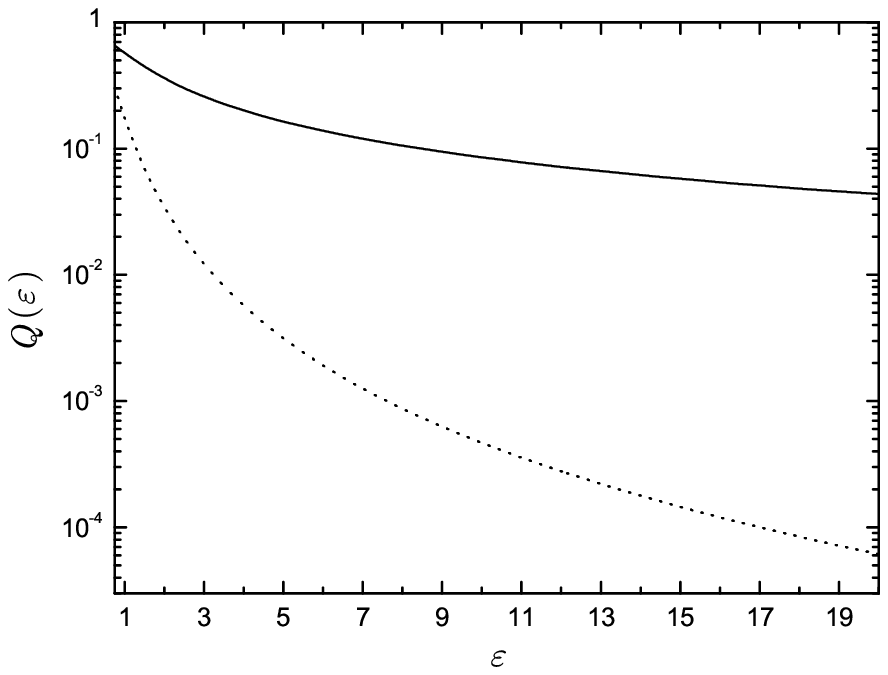}}
\caption{\label{fig13} The universal scaling $Q(\varepsilon)$ is
presented as a function of the dimensionless energy $\varepsilon$ of
the incident electron. Solid line, excitation into the singlet
$2^1S$ state; dotted line, excitation into the triplet $2^3S$ state.
The calculations are performed according to Eqs.~\eqref{eq24},
\eqref{eq27}, and \eqref{eq29}.}
\end{figure}

Another situation occurs, if one can experimentally distinguish the
excitations into the singlet $2^1S$ and the triplet $2^3S$ states of
helium-like ion. The corresponding cross sections,
$d\sigma_{\mathrm{s}}^*(\mathrm{He})$ and
$d\sigma_{\mathrm{t}}^*(\mathrm{He})$, are given by the formulae
similar to Eqs.~\eqref{eq23} and \eqref{eq24}, where the
dimensionless function $T$ is given by
\begin{equation}
T = \begin{cases} \label{eq29}
\dfrac{2}{4}|2 M_{\mathrm{a}}- M_{\mathrm{b}}|^{2}, &
\text{singlet $2^1S$ state}\,  ;\\
2 \dfrac{3}{4}|M_{\mathrm{b}}|^{2}, & \text{triplet $2^3S$ state}\,
.
\end{cases}
\end{equation}
Here the factor 2 accounts for the number of electrons in a
helium-like ion. The factors $1/4$ and $3/4$ are the statistical
weights for the singlet and triplet states, respectively. The
dimensionless functions $M_{\mathrm{a}}$ and $M_{\mathrm{b}}$ are
given by Eqs.~\eqref{eq13} and \eqref{eq14}. As it is seen, the
excitation of the triplet $2^3S$ state occurs only due to the
exchange interaction. In Figs.~\ref{fig9} and \ref{fig10}, we
present the universal functions $F(\varepsilon,\theta)$ describing
the energy and angular behavior of the differential cross sections
for excitation of helium-like ion into the $2^1S$ and the $2^3S$
states, respectively. In Figs.~\ref{fig11} and \ref{fig12}, the same
functions $F(\varepsilon,\theta)$ are given with respect to the
dimensionless energy for the forward ($\theta=0$) and backward
($\theta=\pi$) scattering, respectively. Integrating the functions
$F(\varepsilon,\theta)$ over the solid angle yields the universal
functions $Q(\varepsilon)$, which define the energy dependence of
the total cross sections $\sigma_{\mathrm{s,t}}^*(\mathrm{He})$ (see
Fig.~\ref{fig13}). The latter obey universal scalings similar to
that given by Eq.~\eqref{eq26}.

The averaged cross section for excitation of helium-like ion into
the $1s2s$ configuration can be written as
\begin{equation}
\label{eq30} d\sigma^*(\mathrm{He}) =
d\sigma_{\mathrm{s}}^*(\mathrm{He}) +
d\sigma_{\mathrm{t}}^*(\mathrm{He}) \,  ,
\end{equation}
which is consistent with Eq.~\eqref{eq28}. A similar relation holds
true also for the total cross sections.

Concluding, we have studied the inelastic electron scattering on
hydrogen- and helium-like ions in the ground state followed by the
excitation of a K-shell electron into the $2s$ state. The universal
scaling behavior for the differential and total cross sections is
deduced. As a method, the consistent non-relativistic perturbation
theory is employed. Since the Feynman diagrams are calculated on the
level of one-photon exchange, our results are valid for multicharged
ions with moderate values of the nuclear charge $Z$.

\acknowledgments

The authors acknowledge I.I.~Tupitsyn for valuable discussions. This
research has received financial support from DFG, BMBF, GSI, and
INTAS (Grant no. 06-1000012-8881).



\begin{thebibliography}{99}
\bibitem{1} R.J.W. Henry, Phys. Rep. 68 (1981) 1.
\bibitem{2} Y. Itikawa, Phys. Rep. 143 (1986) 69.
\bibitem{3} Y. Itikawa, Atomic Data Nucl. Data Tables 63 (1996) 315.
\bibitem{4} V.I. Fisher, Y.V. Ralchenko, V.A. Bernshtam, A. Goldgirsh,
Y. Maron, L.A. Vainshtein, I. Bray, H. Golten, Phys. Rev.
A 55 (1997) 329.
\bibitem{5} Y.-D. Jung, Astrophys. J. 396 (1992) 725.
\bibitem{6} L.D. Landau, E.M. Lifshits, Quantum Mechanics: Non-relativistic
Theory, third ed., Pergamon Press, London, 1977.
\bibitem{7} A.I. Mikhailov, I.A. Mikhailov, A.N. Moskalev,
A.V. Nefiodov, G. Plunien, G. Soff, Phys. Rev. A 69 (2004) 032703.
\bibitem{8} K.T. Dolder, B. Peart, J. Phys. B 6 (1973) 2415.
\end{thebibliography}
\end{document}